\documentclass[ prd,twocolumn]{revtex4}
\usepackage{xcolor}
\usepackage[utf8]{inputenc}
\usepackage{graphicx}
\usepackage{empheq}
\usepackage{braket}
\usepackage{float}
\usepackage{amssymb}
\usepackage{amsmath}
\usepackage{mhchem}
\usepackage{siunitx}
\usepackage[titletoc]{appendix}

\usepackage{hyperref}
\hypersetup{colorlinks=true, linkcolor=black, citecolor=blue, urlcolor=black}

\begin{document}

\title{Magnetic transitions induced by pressure and magnetic field in a two-orbital $5f$-electron 
model in cubic and tetragonal lattices}
\author{A. C. Lausmann$^{1}$\href{https://orcid.org/0000-0002-2085-8229}{\includegraphics[scale=0.05]{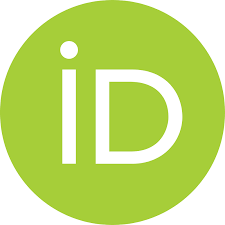}},  E. J. Calegari$^{1}$\href{https://orcid.org/0000-0002-0157-7835}{\includegraphics[scale=0.05]{orcid.png}}}
\author{Julián Faúndez$^{2}$\href{https://orcid.org/0000-0002-6909-0417}{\includegraphics[scale=0.05]{orcid.png}}}
\author{P. S. Riseborough$^{3}$\href{https://orcid.org/0000-0002-2216-3586}{\includegraphics[scale=0.05]{orcid.png}}}
\author{S. G. Magalhaes$^{2}$\href{https://orcid.org/0000-0002-6874-7579}{\includegraphics[scale=0.05]{orcid.png}}}

\affiliation{$^{1}$Departamento de Física - Universidade Federal de Santa Maria, 97105-900, Santa Maria, RS, Brazil}
\affiliation{$^{2}$Condensed Matter Physics Group, Instituto de Física, Universidade Federal do Rio Grande do Sul, $91501-970$ Porto Alegre, RS, Brazil}
\affiliation{$^{3}$Physics Department, Temple University, Philadelphia, Pennsylvania 19122, USA}

\date{\today}
\begin{abstract} 
We investigate the onset and evolution of under the simultaneous application of pressure and magnetic field of
distinct itinerant N\'eel states using the underscreened Anderson Lattice Model (UALM) which has been proposed to describe $5f$-electron systems.
The model is composed by two narrow $f$-bands (of either $\alpha$ or $\beta$ character) that hybridize with a wide $d$-band and local $5f$-electron interactions.
We consider both cubic and tetragonal lattices. The N\'eel order parameters $\phi^{\beta}$ and $\phi^{\alpha}$ are assumed to be fixed by an Ising anisotropy. The applied magnetic field $h_z$ is parallel to the anisotropy axis. It has been assumed that the variation of the band width $W$ is sensitive to pressure. In the absence of a magnetic field, the increase of $W$ takes the system from the phase AF$_1$ to another phase AF$_2$.  The phase AF$_1$ occurs when $\phi^{\beta}>\phi^{\alpha}>0$
while in the AF$_2$ phase the gaps satisfy $\phi^{\alpha}>\phi^{\beta}>0$.
In the presence of a magnetic field $h_z$, the phase AF$_2$ is quickly suppressed
and reappears again at intermediate values of the magnetic field while it is predominant at higher magnetic fields.  The analysis of the partial density of states close to the phase transition between the phases AF$_1$ and AF$_2$, allows a better understanding the mechanism responsible whereby the transition is induced by an increase in the magnetic field. As a important general result, we found that the magnetic field $h_z$ favours the phase AF$_2$ while the phase AF$_1$ is suppressed.
For the tetragonal lattice, the phase AF$_2$ is even more favored when $h_z$ and $c/a$ increases concomitantly, where $c$ and $a$ are the lattice parameters.
\end{abstract}
\maketitle
\section{Introduction}
\label{Intro}
Electrons $5$\textit{f} are source of intriguing physics due to their dual localized-delocalized character \cite{moore}. This can have profound implications for their collective behavior.
Indeed, the uranium $5$\textit{f}-electrons systems shows a variety of ground states which includes
localized and itinerant magnetism \cite{Santini,Miyake}, 
unconventional superconductivity \cite{Pfleiderer,Aoki}
and the ignematic exotic Hidden Order in  URu$_2$Si$_2$
\cite{Mydosh2011,Oppeneer2014,Wolowiec,Calegari2017}.
Moreover, it is well known that these states
can be tuned by pressure (hydrostatic or chemical) and magnetic field \cite{Magalhaes2020,Faundez2021}.
This imposes a requirement on any comprehensive microscopic model for $5$\textit{f}-electrons that it should describe the appearance and evolution of quantum conventional, unconventional or even exotic collective states \cite{Valiska2018, Correa2012, Niklowitz210}.

The Underscreened Anderson Lattice Model (UALM) has been introduced as a generalization of the Underscreened Kondo Lattice
which successfully described certain aspects of the physics of uranium compounds \cite{Riseborough2012, Perkins, Thomas}. In particular, 
the interplay of the Kondo effect and ferromagnetism \cite{Bernard}. The UALM includes the direct hopping between distinct orbitals ($\chi=\alpha $ and $\beta$) which gives rise to two quite narrow $f$-bands. These $f$-bands in turn, are hybridized with a wide $d$-band.  Lastly, there are $f$-electron intra- and inter-orbitals interactions. Remarkably,  this model can host itinerant spins orderings in which time reversal symmetry may or may not be broken \cite{Shah2000,Yuan1}.

In the present work, we study how conventional Spin Density Waves (SDWs) develop in the UALM under simultaneous application of pressure and magnetic fields for both cubic and tetragonal lattices. The applied pressure is mimicked by an increase in the bandwidths.  We also assume an Ising-like anisotropy for the magnetic order parameters (OPs) \cite{Schoenes1981,Havela1992,Maskova2019}. The magnetic field is applied along the axis of anisotropy. Even for the conventional spin ordering,
 the two-orbital nature of model has an unusual role. In particular, the SDW  exist in two distinct phase (with same nesting vectors) AF$_1$ and AF$_2$ that are characterized by the relative magnitudes of the orbital's staggered magnetizations \cite{Magalhaes2020,Faundez2021}. For low pressure, there is the onset of AF$_1$ at lower temperature. When the pressure is increased,  AF$_2$ starts to compete by the stability with AF$_1$.
Eventually, the initial AF$_1$ is abruptly replaced by AF$_2$.

The SDWs also present noteworthy features concerning their metallic characters.
For instance, it is possible that the Fermi energy, can be located inside the two spin gaps or even inside one spin gap but not inside the other.
However, with the magnetic field aligned with the Ising direction, the Zeeman splitting between the spin-up and spin-down sub-bands is strongly affected as the field changes.
As consequence, the level of metallicity can be also affected.
Therefore, an evolution of the Fermi Surface (FS) can be anticipated with some type of reconstruction as the combined application of pressure and magnetic field changes.

We also highlight the possibility of metamagnetic transitions \cite{Miyake,Ken, Solhanek, Mushnikov}. Since, the two SDWs compete for stability when the pressure is increased, magnetic field modifications of the band structure may produce transitions \cite{Bruck1994,Pospisil2018}.
In such cases, one may expect that the induced transitions will be highly sensitive to the lattice structure, such as cubic or tetragonal.

This paper is organized as follow: the model is presented in the section \ref{model_1}. Then, we derive the Green's functions, gap equations and free energy which are presented in section \ref{FGreen}. In section \ref{Results}, we present the phase diagrams for pressure and magnetic fields for cubic and tetragonal lattices.
The conclusions and other remarks are found in section \ref{Conclusion}.

\section{Model}
\label{model_1}

In this work we investigate the effects of an applied magnetic field and pressure in an UALM Hamiltonian for cubic and tetragonal lattices. The Hamiltonian is written as follow
\begin{equation}
   \hat{H} = \hat{H}_f + \hat{H}_d + \hat{H}_{fd}\, .
  \label{hamilt}
\end{equation}

The first term $\hat{H}_f$
represents the $5$\textit{f}-electron and is composed of two parts:
\begin{equation}
\hat{H}_f=\hat{H}_{f,0}+ \hat{H}_{f,int}\,,
\label{int}
\end{equation}
where the noninteracting part $H_ {0, f}$ describes two degenerate narrow $5$\textit{f} bands given as below
\begin{equation}
\hat{H}_{f,0} = \sum_{\vec{k},\sigma}\sum_{\chi} \ E_{f}^{\chi}(\vec{k}) \
f^{\dag \chi}_{\vec{k},\sigma}f^{\chi}_{\vec{k},\sigma}\,.
\label{eq0}
\end{equation}
The $f^{\dag \chi}_{\vec{k},\sigma}$ are the creation operators for electrons with
spin $\sigma(=\uparrow,\downarrow )$ at site $i$. The $\chi$-bands ($\chi=\alpha$ and $\beta$) in Eq. (\ref{eq0}) follow the intraband and interband nesting property $E^{\chi}_{f}(\vec{k}+\vec{Q})=-E^{\chi'}_{f}(\vec{k})$
where $\chi=\chi'$  or $\chi\neq\chi'$. The electron band dispersion relations are described by
$E_{f}^{\chi}(\vec{k})$ and the vector $\vec{Q}$ is a commensurate momentum transfer in the first Brillouin zone.

The second term in Eq. (\ref{int}) represents the local Coulomb and Hund's rule exchange interactions, which are given by
\begin{align}
\hat{H}_{f, int} \ = \ \bigg(\frac{U - J}{2 N}\bigg) \
\sum_{\vec{k},\vec{k}',\vec{q},\sigma}\sum_{\chi\ne\chi'}
\ f^{\dag,\chi}_{\vec{k}+\vec{q},\sigma} \ f^{\chi}_{\vec{k},\sigma} \
f^{\dag,\chi'}_{\vec{k}'- \vec{q},\sigma} f^{\chi'}_{\vec{k}',\sigma}
\nonumber \\
+ \ \bigg({ U \over 2 N}\bigg)
\sum_{\vec{k},\vec{k}',\vec{q},\sigma}\sum_{\chi,\chi'}
\ f^{\dag,\chi}_{\vec{k}+\vec{q},\sigma} \ f^{\chi}_{\vec{k},\sigma} \
f^{\dag,\chi'}_{\vec{k}'-\vec{q},-\sigma} f^{\chi'}_{\vec{k}',-\sigma}\nonumber
 \\ +  \bigg({J \over 2 N}\bigg)
\sum_{\vec{k},\vec{k}',\vec{q},\sigma}\sum_{\chi\ne\chi'} \
f^{\dag,\chi}_{\vec{k}+\vec{q},\sigma} \  f^{\chi'}_{\vec{k},\sigma} \
f^{\dag,\chi'}_{\vec{k}'-\vec{q},-\sigma} \ f^{\chi}_{\vec{k}',-\sigma}\,.
\label{Hfintreal}
\end{align}

The second term  in Eq. (\ref{hamilt})  represents the conduction electron term $\hat{H}_d$
\begin{equation}
    \hat{H}_d \ = \ \sum_{\vec{k},\sigma} \ \epsilon_d(\vec{k}) \ d^{\dag}_{\vec{k},\sigma} \ d_{\vec{k},\sigma},
\label{Hd}
\end{equation}
where $\epsilon(\vec{k})$ describes the dispersion relation of
conduction electrons labeled by the Bloch wave vector $\vec{k}$.
The term $\hat{H}_{fd}$
describes the on-site hybridization process in the considered model by
{\small
\begin{equation}
\hat{H}_{fd} =  \sum_{\vec{k},\sigma}\sum_{\chi=\alpha\beta} \ \bigg(V_{\chi}(\vec{k})
\ f^{\dag \chi}_{\vec{k},\sigma}  d_{\vec{k},\sigma}  +
V^*_{\chi}(\vec{k}) \ d^{\dag}_{k,\sigma}  f^{\chi}_{\vec{k},\sigma} \ \bigg)\,.
\label{Hhyb}
\end{equation}}
The effects of an applied magnetic field are taken into account by inclusion of an additional term in Eq. (\ref{hamilt}) given by
\begin{equation}
\hat{H}_{ext}=-  \sum_{\vec{k}} \sum_{\sigma=\pm}  \sigma  [ H_{z} ^{f}f^{\dag}_{ \vec{k},\sigma}  f_{\vec{k},\sigma}+ H_{z}^{d}  d^{\dag}_{\vec{k},\sigma}  d_{\vec{k},\sigma}]
\label{campo}
\end{equation}
with 
\begin{equation}
H_z^{f(d)}= g_{f(d)}\mu_B h_z.
\label{eqHz}
\end{equation}
The value $\sigma=1$ and $-1$ correspond to the up and down spin projections, respectively.

The dispersion relations for the $5f$ band, $E_f^{\chi}(\vec{k})=\epsilon_{f}+\epsilon_f(\vec{k})$, and the conduction band, $\epsilon_d(\vec{k})$,  refer to a simple tetragonal
lattice. Thus
\begin{equation}
\epsilon_{A}(\vec{k})=-2t_{A,a}[\cos(k_xa)+\cos(k_ya)]-2t_{A,c}\cos(k_zc)
\label{Ek}
\end{equation}
in which $A=f$ or $d$, and $a$ and $c$ are the lattice parameters. If $a=c$, we have a cubic lattice.


\section{Green's functions, Gaps and Free energy}
\label{FGreen}

The temporal and spatial Fourier transform of one-particle $f$-electron Green's function
satisfy the equations of motion
given by:
\begin{align}
[ \ \omega \ - \ \tilde{E}_{f,\sigma}^{\alpha}(\vec{k}) \ ] \ \langle\langle f^{\alpha}_{\vec{k},\sigma};f^{\dag\chi'}_{ \vec{k'},\sigma} \rangle\rangle_{\omega} =\delta^{\alpha,\chi'}\ \delta_{\vec{k},\vec{k}'}+ V_{\alpha}(\vec{k})\times  \nonumber\\
    \langle\langle d_{\vec{k},\sigma};f^{\dag\chi'}_{ \vec{k'},\sigma} \rangle\rangle_{\omega}
+\bigg({{U-J} \over 2}\bigg)
[\langle\langle\hat{\rho}^{\beta}_{-\vec{Q},\sigma} f^{\alpha}_{\vec{k}-\vec{Q},\sigma};f^{\dag\chi'}_{ \vec{k'},\sigma} \rangle\rangle_{\omega}   \nonumber\\+\langle\langle\hat{\rho}^{\beta}_{\vec{Q},\sigma} f^{\alpha}_{\vec{k}+\vec{Q},\sigma};f^{\dag\chi'}_{ \vec{k'},\sigma} \rangle\rangle_{\omega}]\nonumber \\ +\bigg({U \over 2 }\bigg)
[\sum_{\chi''}\langle\langle\hat{\rho}^{\chi''}_{-\vec{Q},-\sigma} f^{\alpha}_{\vec{k}-\vec{Q},\sigma};f^{\dag\chi'}_{ \vec{k'},\sigma} \rangle\rangle_{\omega}   \nonumber\\+\sum_{\chi}\langle\langle\hat{\rho}^{\chi}_{\vec{Q},-\sigma} f^{\alpha}_{\vec{k}+\vec{Q},\sigma};f^{\dag\chi'}_{ \vec{k'},\sigma} \rangle\rangle_{\omega}]&\
\label{Galpha}
\end{align}
and

\begin{align}
[ \ \omega \ - \ \tilde{E}_{f,\sigma}^{\beta}(\vec{k}) \ ] \ \langle\langle f^{\beta}_{\vec{k},\sigma};f^{\dag\chi'}_{ \vec{k'},\sigma} \rangle\rangle_{\omega} = \delta^{\beta,\chi'}  \ \delta_{\vec{k},\vec{k}'}+ V_{\beta}(\vec{k}) \times \nonumber\\
 \  \langle\langle d_{\vec{k},\sigma};f^{\dag\chi'}_{ \vec{k'},\sigma} \rangle\rangle_{\omega}\
+\bigg(\frac{U-J}{2}\bigg)
[\langle\langle\hat{\rho}^{\alpha}_{-\vec{Q},\sigma} f^{\beta}_{\vec{k}-\vec{Q},\sigma};f^{\dag\chi'}_{ \vec{k'},\sigma} \rangle\rangle_{\omega}   \nonumber\\+\langle\langle\hat{\rho}^{\alpha}_{\vec{Q},\sigma} f^{\beta}_{\vec{k}+\vec{Q},\sigma};f^{\dag\chi'}_{ \vec{k'},\sigma} \rangle\rangle_{\omega}]\nonumber \nonumber\\  +\bigg(\frac{U}{2}\bigg)
[\sum_{\chi''}\langle\langle\hat{\rho}^{\chi''}_{-\vec{Q},-\sigma} f^{\beta}_{\vec{k}-\vec{Q},\sigma};f^{\dag\chi'}_{ \vec{k'},\sigma} \rangle\rangle_{\omega}  \nonumber\\+\sum_{\chi}\langle\langle\hat{\rho}^{\chi}_{\vec{Q},-\sigma} f^{\beta}_{\vec{k}+\vec{Q},\sigma};f^{\dag\chi'}_{ \vec{k'},\sigma} \rangle\rangle_{\omega}]&\
\label{Gbeta}
\end{align}
where $\tilde{E}_{f,\sigma}^{\chi}(\vec{k})=E_f^{\chi}(\vec{k}) -\sigma H_{z}^{f}$ and where
\begin{equation}
  \hat{\rho}^{\chi}_{\vec{Q},\sigma} = \bigg(\frac{1}{ N}\bigg) \sum_{\vec{k}'} f^{\dag,\chi}_{\vec{k}'+\vec{Q},\sigma}  f^{\chi}_{\vec{k}',\sigma}\,.
\end{equation}

We use the Hartree-Fock approximation to decouple
the two-particle Green's functions terms in Eqs. (\ref{Galpha})-(\ref{Gbeta}). Thus
\begin{equation}
\langle\langle \hat{\rho}^{\chi}_{\vec{Q},\sigma} f^{\chi'}_{\vec{k}-\vec{Q},\sigma};f^{\dag\chi''}_{ \vec{k'},\sigma} \rangle\rangle_{\omega} \simeq n^{\chi}_{\vec{Q},\sigma}\langle\langle  f^{\chi'}_{\vec{k}-\vec{Q},\sigma};f^{\dag\chi''}_{ \vec{k'},\sigma} \rangle\rangle_{\omega}\,,
\label{aprox1}
\end{equation}
where $\vec{Q}$ is the nesting vector.
Therefore, the Green's function equations of motion become:
\begin{align}
&&[ \ \omega \ - \ E_{f,\sigma}^{\alpha}(\vec{k}) \ ] \ \langle\langle f^{\alpha}_{\vec{k},\sigma};f^{\dag\chi'}_{ \vec{k'},\sigma} \rangle\rangle_{\omega} =\delta^{\alpha,\chi'}  \ \delta_{\vec{k},\vec{k}'}
 \nonumber\\&& +V_{\alpha}(\vec{k})   \langle\langle d_{\vec{k},\sigma};f^{\dag\chi'}_{ \vec{k'},\sigma} \rangle\rangle_{\omega}\
-\phi^{\alpha}_{\sigma}  \langle\langle f_{\vec{k}+\vec{Q},\sigma}^{\alpha};f^{\dag\chi'}_{ \vec{k'},\sigma} \rangle\rangle_{\omega}\
\label{Galpha_apHF}
\end{align}
and
\begin{align}
&&[ \ \omega \ - \ E_{f,\sigma}^{\beta}(\vec{k}) \ ] \ \langle\langle f^{\beta}_{\vec{k},\sigma};f^{\dag\chi'}_{ \vec{k'},\sigma} \rangle\rangle_{\omega} =\delta^{\beta,\chi'}  \ \delta_{\vec{k},\vec{k}'}
 \nonumber\\&& +V_{\beta}(\vec{k})   \langle\langle d_{\vec{k},\sigma};f^{\dag\chi'}_{ \vec{k'},\sigma} \rangle\rangle_{\omega}\
-\phi^{\beta}_{\sigma}  \langle\langle f_{\vec{k}+\vec{Q},\sigma}^{\beta};f^{\dag\chi'}_{ \vec{k'},\sigma} \rangle\rangle_{\omega}\,,
\label{Gbeta_apHF}
\end{align}
where the
spin-dependent Hartree-Fock dispersion relation in Eqs. (\ref{Galpha_apHF})-(\ref{Gbeta_apHF})
is given by
\begin{equation}
    E_{f, \sigma}^{\chi}(\vec{k})=
\sum_{\chi'} \bigg( U  n^{\chi'}_{-\sigma}+(U - J)  n^{\chi'}_{\sigma}
     (1-\delta^{\chi,\chi'}) \bigg)
     +\tilde{E}_{f,\sigma}^{\chi}(\vec{k})
\label{Erenorm}
\end{equation}
and the spin gap for the $\chi$-orbital is
\begin{equation}
\phi^{\chi}_{\sigma}
= \sum_{\chi'} (U \ n^{\chi'}_{\vec{Q},-\sigma} +
 (U - J) \ n^{\chi'}_{\vec{Q},\sigma} (1-\delta^{\chi,\chi'}))\,.
\label{GapNeel}
\end{equation}

The mixed $f-d$ Green's function satisfies the following equation
\begin{align}
&&     [ \ \omega \ - \ \epsilon(\vec{k}) \ ] \ \langle\langle d_{\vec{k},\sigma};f^{\dag\chi'}_{\vec{k'},\sigma}\rangle\rangle_{\omega} \ =
 \ V_{\alpha}(\vec{k})^* \
\langle\langle f^{\alpha}_{\vec{k},\sigma};f^{\dag\chi'}_{\vec{k'},\sigma}\rangle\rangle_{\omega}+ \nonumber\\
&&\ V_{\beta}(\vec{k})^* \
\langle\langle f^{\beta}_{\vec{k},\sigma};f^{\dag\chi'}_{\vec{k'},\sigma}\rangle\rangle_{\omega}\,.
\label{Gmix}
\end{align}
We will choose a basis set for the $f$ orbitals, such that $V_{\beta}({\vec{k}})=0$ and $V_{\alpha}({\vec{k}})=V_{\alpha}$
simply to avoid the transformation to a new basis set. The choice of basis states should not change the main physical results, as discussed in ref. \cite{Riseborough2012}. Thus, we obtain the Green’s functions that can be recast in the
form
\begin{equation}
\langle\langle f_{\vec{k}+\vec{Q},\sigma}^{\beta};f^{\dag\beta}_{ \vec{k'},\sigma} \rangle\rangle_{\omega}=\sum_{\pm}
\frac{Z^{\beta}_{\pm,\vec{k}}}{\omega-\omega^{\beta}_{\pm,\vec{k}}}
\label{Gb}
\end{equation}
and
\begin{equation}
\langle\langle f_{\vec{k}+\vec{Q},\sigma}^{\alpha};f^{\dag\alpha}_{ \vec{k'},\sigma} \rangle\rangle_{\omega}=\sum_{i=1}^{4}
\frac{Z^{\alpha}_{i,\vec{k}}}{\omega-\omega^{\alpha}_{i,\vec{k}}}\,,
\label{Ga}
\end{equation}
where the weights for the $\beta$ character are given by
\begin{equation}
Z^{\beta}_{\pm,\vec{k}}=\frac{\pm \phi^{\beta}_{\sigma} }{\omega^{\beta}_{+,\vec{k}}-\omega^{\beta}_{-,\vec{k}}}
\end{equation}
and $\omega^{\beta}_{\pm,\vec{k}}$ are the dispersion relations of the bands given by the zeros of the denominator
$\varphi_{\vec{Q}}^{\beta}(\omega)$ of the Green's function \newline$\langle\langle f_{\vec{k}+\vec{Q},\sigma}^{\beta};f^{\dag\beta}_{ \vec{k'},\sigma} \rangle\rangle_{\omega}$ :
\begin{equation}
\varphi_{\vec{Q}}^{\beta}(\omega)=(\omega-E_{f,\sigma}^{\alpha}(\vec{k}))
(\omega-E_{f,\sigma}^{\alpha}(\vec{k}+\vec{Q}))-(\phi^{\beta}_{\sigma})^{2}\,.
\label{Dbeta}
\end{equation}
Also, the weights for the $\alpha$ character are expressed by
\begin{equation}
Z^{\alpha}_{1,3,\vec{k}}=\pm\frac{[(\omega^{\alpha}_{1,\vec{k}})^{2}-(\epsilon(\vec{k}))^{2}]\phi^{\alpha}_{\sigma}}{2\omega^{\alpha}_{1,\vec{k}}[(\omega^{\alpha}_{1,\vec{k}})^{2}-(\omega^{\alpha}_{2,\vec{k}})^{2}]}\
\end{equation}
and
\begin{equation}
Z^{\alpha}_{2,4,\vec{k}}=\mp\frac{[(\omega^{\alpha}_{2,\vec{k}})^{2}-(\epsilon(\vec{k}))^{2}]\phi^{\alpha}_{\sigma}}{2\omega^{\alpha}_{2,\vec{k}}[(\omega^{\alpha}_{1,\vec{k}})^{2}-(\omega^{\alpha}_{2,\vec{k}})^{2}]}.
\end{equation}
Similarly to the $\beta$ character case, the  dispersion  relations $\omega^{\alpha}_{i,\vec{k}}$ of  the  $\alpha$ character bands are  given  by the  zeros  of  the  denominator  $\varphi_{\vec{Q}}^{\alpha}(\omega)$ of the Green's function $\langle\langle f_{\vec{k}+\vec{Q},\sigma}^{\alpha};f^{\dag\alpha}_{ \vec{k'},\sigma} \rangle\rangle_{\omega}$  :
\begin{align}
\label{Dalpha}
\varphi_{\vec{Q}}^{\alpha}(\omega)&=&-(\phi^{\alpha}_{\sigma})^{2}(\omega-\epsilon(\vec{k}))(\omega-\epsilon(\vec{k}+\vec{Q}))\nonumber\\&&+[(\omega-E_{f,\sigma}^{\alpha}(\vec{k}))(\omega-\epsilon(\vec{k}))-|V_{\alpha}(\vec{k})|^{2}]\times \nonumber\\&&
[(\omega-E_{f,\sigma}^{\alpha}(\vec{k}+\vec{Q}))(\omega-\epsilon(\vec{k}+\vec{Q}))-|V_{\alpha}(\vec{k})|^{2}]\,.
\end{align}

The Green's function given in Eqs. (\ref{Gb})-(\ref{Ga}) form a closed set of equations, which can be solved exactly.
Thus, the spin gap in Eq. (\ref{GapNeel}) can be calculated directly from
\begin{equation}
n^{\chi}_{\vec{Q},\sigma}=
\frac{1}{N} \sum_{\vec{k}}
 \oint \frac{d\omega}{2\pi i} f(\omega) \
 \langle\langle f_{\vec{k}+\vec{Q},\sigma}^{\chi};f^{\dag\chi}_{ \vec{k'},\sigma} \rangle\rangle_{\omega}\,. 
\label{GF}
\end{equation}

In Eq. (\ref{GF}), the contour of the path integral encircles the real axis without enclosing any poles  of the Fermi-Dirac distribution $f(\omega)$.
Therefore, we can explore
a scenario where
the instability of the paramagnetic phase towards to two distinct
phases
SDWs occurs at the same nesting vector $\vec{Q}$ given by the spin gaps in distinct orbitals.
However, it is important to be noted that  the spin gaps $\phi^{\alpha}_{\sigma}$ and $\phi^{\beta}_{\sigma}$ are, indeed, coupled.
Moreover, the spin gaps are proportional to the magnetic  order  parameters and present exactly same behavior. 

In the Hartree-Fock approximation, since $\phi^{\alpha(\beta)}_{\uparrow\downarrow}=\mp\phi^{\alpha(\beta)}$, the free energy can be expressed in terms of the gaps by:
\begin{eqnarray}
f_{HF}&=&\Omega(T,\mu)+ \mu N_{tot} +\frac{N}{U^2-J^2} \  [U\left((\phi^{\alpha})^2 + (\phi^{\beta})^2\right)\nonumber\\&& -2J\phi^{\alpha}\phi^{\beta}]\,,
    \label{freeenergy}
\end{eqnarray}
in which $\mu$ is the chemical potential,  $N_{tot}=n_f^{\alpha}+n_f^{\beta}+n_d$ ($n_d$ is the average occupation of the conduction electrons) and
\begin{equation}
 \Omega(T,\mu)= -k_{B} T \frac{1}{N}\sum_{\vec{k}}\sum_{\gamma} \ln[1+ e^{-\frac{(E^{\gamma}-\mu)}{k_{B} T}}],
     \label{freeenergy0}
\end{equation}
where $N$ is the number of sites in the lattice. The quantity $E^{\gamma}$ represents the dispersion relation of the $\beta$ and $\alpha$ bands that are defined by the conditions $\varphi_{\vec{Q}}^{\beta}(\omega)=0$ and $\varphi_{\vec{Q}}^{\alpha}(\omega)=0$, in Eqs. (\ref{Dbeta}) and (\ref{Dalpha}), respectively.




\section{Numerical results}\label{Results}

The numerical results presented in this section were calculated
for $J=5U$ with $U=0.165$ eV  and the total occupancy $N_{tot} = 1.609$. This choice for $N_{tot}$ places the $5f$ bands close to half-filling, which favors the instability of the paramagnetic state with respect to Néel antiferromagnetism. It is important to make clear that the above parameters were chosen in order to emphasize the existence of the competing AF phases. We did not intend to describe any particular compound.

\begin{figure}[!htb]
    \centering
    \includegraphics[scale=1.,angle=0]{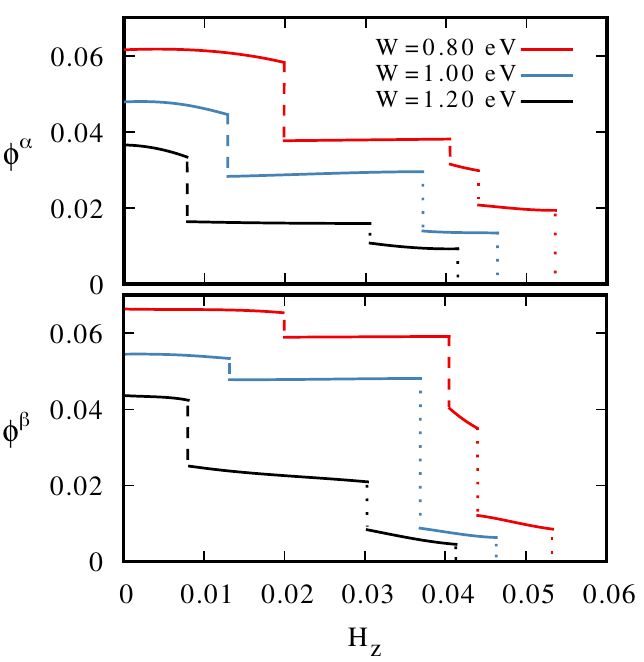}
    \caption{The Neel gaps $\phi^{\alpha}$ and $\phi^{\beta}$
    for the simple cubic lattice as a function of 
    $H_z$ for zero temperature and different values for the band width $W$. The dotted lines denote the AF$_{1}$-AF$_{2}$ and AF$_{2}$-PM first-order transitions while the dashed lines are associated with metamagnetic-like transitions.  }
    \label{POHZC}
\end{figure}
\begin{figure}[!bh]
    \centering
    \includegraphics[scale=01.0,angle=0]{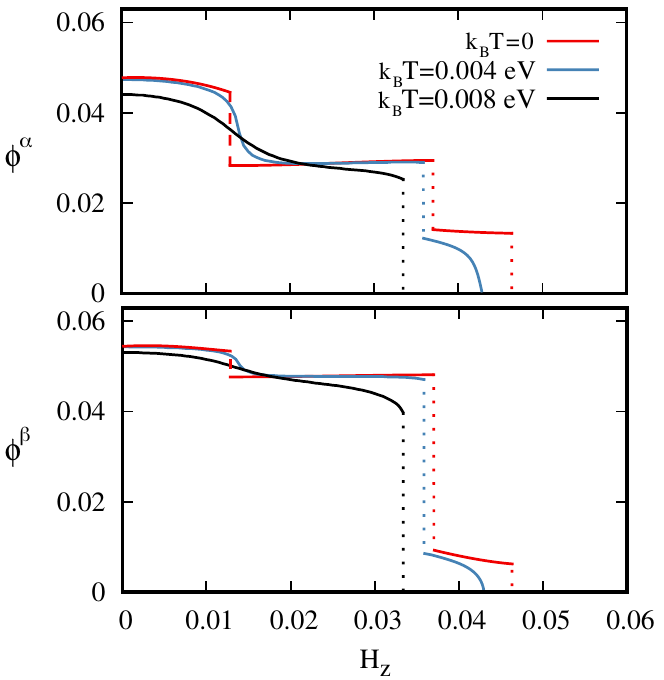}
    \caption{The Neel gaps $\phi^{\alpha}$ and $\phi^{\beta}$
    for the simple cubic lattice as a function of the magnetic field $h_z$ for $W=1.00$ eV and different values of temperature. The dotted and the dashed lines play the same role as in Fig. \ref{POHZC}.}
    \label{POHZCWfixo}
\end{figure}

\begin{figure}[!th]
    \centering
    \includegraphics[scale=0.7,angle=0]{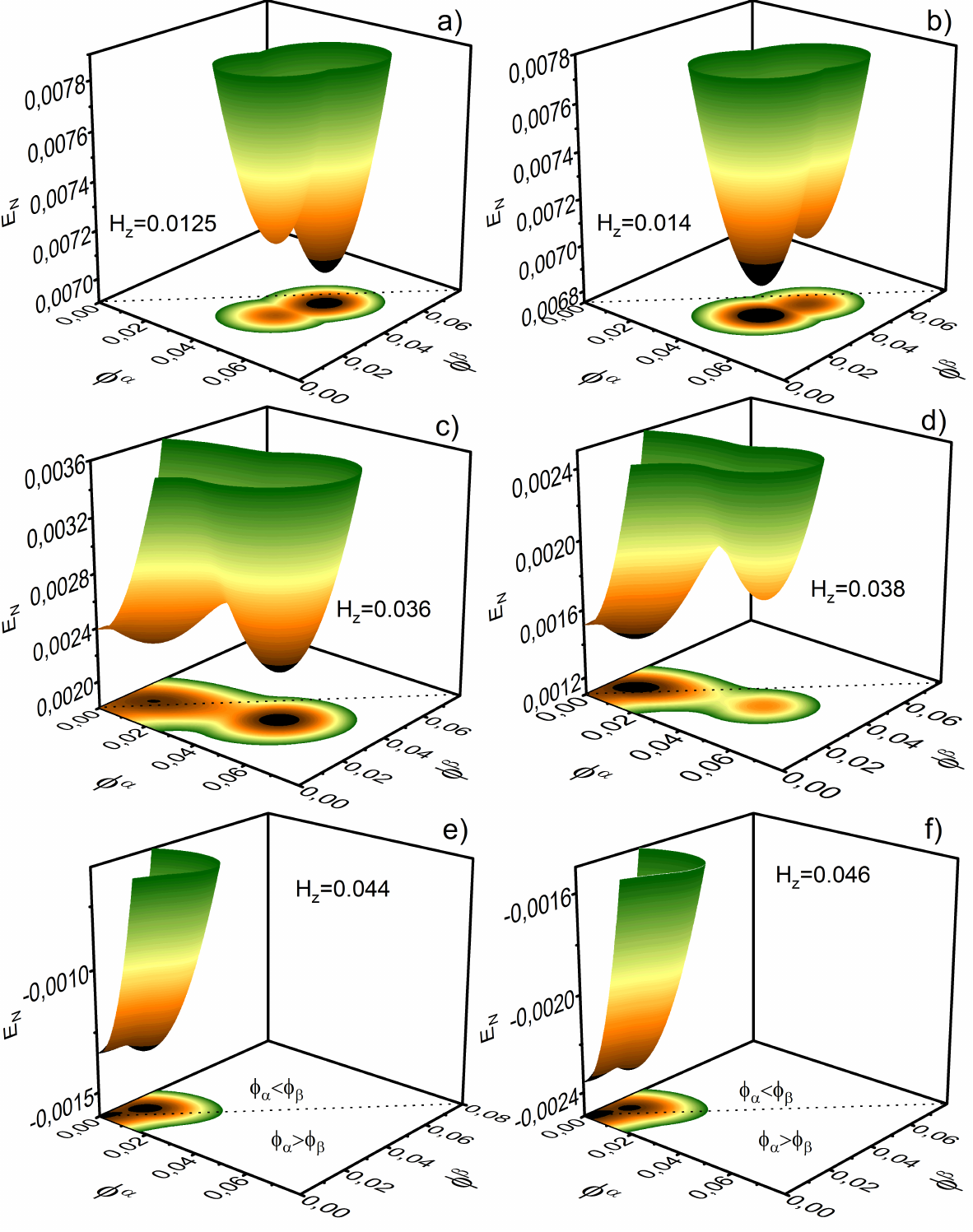}
    \caption{The energy per unit cell at zero temperature. The
projection of the ground state energy on the plan of the gaps show dark regions, in which the self-consistent solutions for the gaps
$\phi^{\alpha}$ and $\phi^{\beta}$ are to be found. The diagonal dotted line is defined by the condition $\phi^{\alpha}=\phi^{\beta}$.}
    \label{FNRC}
\end{figure}

The value of the effective $5f$ level energy was chosen to be $\epsilon_{f} = 0.3$ eV. The width of the conduction ($5f$) band is defined as $2W_{d(f)}$ with $W_{f}/W_{d}=0.3$. It has been assumed that the bandwidths $W_{d(f)}$ are sensitive to external pressure.
We also considered a $\vec{k}$-independent hybridization $V_{\alpha}(\vec{k})=V_{\alpha}= W_{d}/10$. With the purpose to simplify the notation, from now on, we use $W_{d} = W$.

\subsection{Simple Cubic Lattice}
Firstly, we present results for a simple cubic lattice, for which $a=c$ and $t_{A,c}=t_{A,a}$, in the dispersion relation given in Eq. \ref{Ek}.

\begin{figure*}[!t]
    \centering
    \includegraphics[scale=1.0,angle=-90]{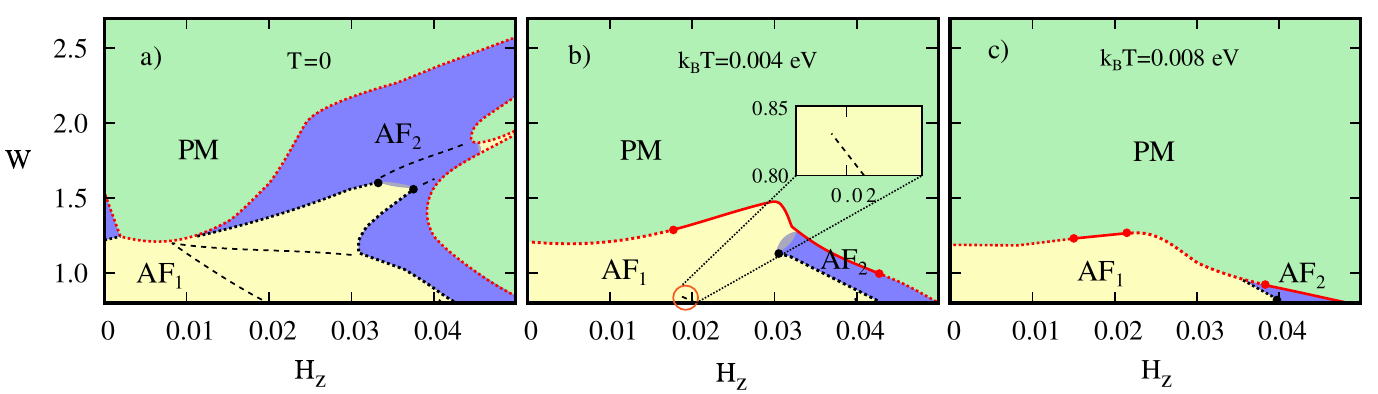}
    \caption{The  phase diagram  for a simple cubic lattice with the band width $W$ versus $H_z$ for several temperatures. The solid and the dotted lines denote second-order and first-order
transition, respectively. The dashed lines mark metamagnetic-like transitions.}
    \label{WHZC}
\end{figure*}

\begin{figure}[!hb]
    \centering
    \includegraphics[scale=0.85,angle=-90]{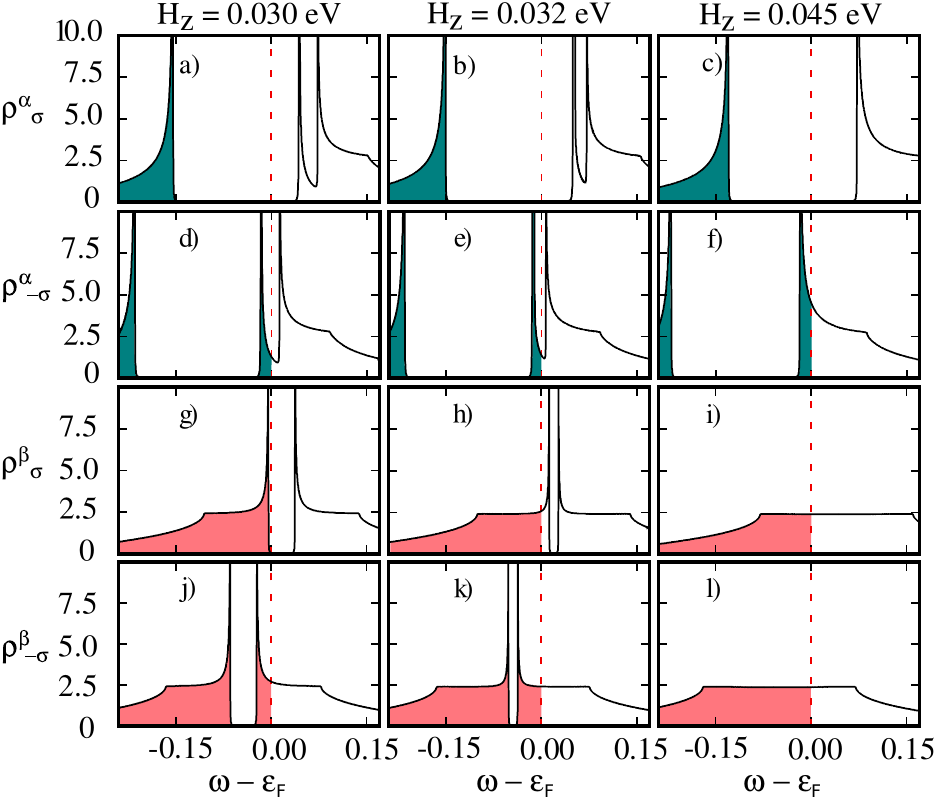}
    \caption{
    The $\alpha$ and $\beta$ partial densities of states for
$W=1.20$ eV, $T=0$, and different
values of $H_z$. The values of $H_z$ have been chosen in order to show the densities of states behavior inside each phase of the diagram presented in the Fig. \ref{WHZC}.}
    \label{DOS_RCWfixo}
\end{figure}

The Neel gaps $\phi^{\alpha}$ and $\phi^{\beta}$
at $T=0$, are shown in Fig. \ref{POHZC} as a function of $H_z$ which is directly proportional to the magnetic field $h_z$ (see Eq. (\ref{eqHz})). Both gaps exhibit discontinuities which indicate the occurrence of first-order phase transitions. The discontinuities denoted by the dotted lines mark first order transitions between two competing antiferromagnetic phases, AF$_{1}$ and AF$_{2}$, or, at higher magnetic fields, between a antiferromagnetic and a paramagnetic phase (PM). The phase AF$_{1}$ is characterized by   $\phi^{\beta}>\phi^{\alpha}>0$, while AF$_{2}$ denotes the phase where $\phi^{\alpha}>\phi^{\beta}>0$. When the system evolves to the PM phase we have $\phi^{\alpha}$= $\phi^{\beta}$= 0. The discontinuities marked by the dashed lines, at lower magnetic field, suggests metamagnetic-like transitions which resemble transitions reported in antiferromagnetic systems \cite{Ken}.

Fig. \ref{POHZCWfixo} displays the Neel gaps
as a function of 
$H_z$, for $W=1.00$ eV and for different temperatures. These results show that the effect of increasing of the temperature is to suppress the discontinues found at low magnetic fields. For $k_BT=0.004$ eV, the transition between the phases AF$_2$ and PM change its nature from first to second order.
On the other hand, the nature of the transition AF$_{1}$-AF$_2$, is unaffected.
Nevertheless, at even higher temperature, for $k_{B}T = 0.008$ eV, the AF$_{1(2)}\rightarrow$ PM phase transition becomes a first-order transition again.
This behavior suggest the existence of tricritical points.

In order to attain better understanding on the discontinuities seen in the Neel gaps shown in Fig. \ref{POHZC} and \ref{POHZCWfixo}, we have analyzed the ground state energy per unit cell E$_{N}$, close to the discontinuities.
The results for $T=0$ and $W=1.00$ eV are shown in Fig. \ref{FNRC} as a function of $\phi^{\alpha}$ and $\phi^{\beta}$. In the figure, we have also included a contour plot of the energy, E$_{N}$. The dark regions are associated to the minima of the ground state energy E$_{N}$, and indicate the places where the self-consistent solutions for the gaps
$\phi^{\beta}$ and $\phi^{\alpha}$, are to be found. A discontinuous transition occurs when the local and the global minima exchange positions.
The dotted line on the contour plot represents the diagonal $\phi^{\alpha}=\phi^{\beta}$. We can distinguish between AF$_1$ - AF$_2$ and the metamagnetic-like transitions based on the position of both local and global minimum relative to the diagonal line.
If both local and global minimum are in the same side of the dotted line when they switch positions, then we have a metamagnetic-like transition. This is the case shown in Figs. \ref{FNRC}(a) and \ref{FNRC}(b). Nevertheless, if the local and the global minimum are on opposite sides of the dotted line, when they switch positions, then there is a first order transition between the AF$_{1}$ and AF$_{2}$ phases, as shown in Figs. \ref{FNRC}(c) and \ref{FNRC}(d).
The first order transition between the phases AF$_2$ and PM is illustrated in Figs. \ref{FNRC}(e) and \ref{FNRC}(f).  In Fig. \ref{FNRC}(e) the global minimum is located in the region where  $\phi^{\alpha}<\phi^{\beta}$ (phase AF$_2$) while in \ref{FNRC}(d) the global minimum is shifted to $\phi^{\alpha}=\phi^{\beta}=0$, characterizing a first order transition to the PM phase. The behavior of $E_{N}$ shown in Fig. \ref{FNRC} is in agreement with the results presented in Fig. \ref{POHZC}.

The effect of the temperature on the boundary of the phases AF$_1$, AF$_2$, and PM, is summarized in the phase diagrams
shown in Fig. \ref{WHZC}.
The dotted lines indicate first order transitions while the solid lines represent second order transitions. The Helmholtz free energy has been used to find the correct positions of the first order transitions on the phase diagrams. In the panel \ref{WHZC}(a), it can be seen that the phase AF$_1$
occurs mainly for low values of $W$ while the AF$_2$ phase is predominant found at higher values of
$H_z$. However, the combination of high values of $W$ and $H_z$, favors the AF$_2$ phase.
For $T=0$, we observe two lines representing transitions between the phases AF$_{1}$ and AF$_{2}$
which end at two critical points localized at the region of $W\approx 1.6$ eV, between $H_z=0.03$ eV and $H_z=0.04$ eV. The critical points are denoted by black solid circles.
For finite temperatures, the AF$_2$ phase is restricted to a small portion of the phase diagram at high magnetic fields and low $W$, as shown in Figs. \ref{WHZC}(b) and \ref{WHZC}(c). On the other hand, the AF$_1$ phase is much more robust to the effect of temperature. With increasing temperature,
there are regions where the first-order AF$_{1(2)}\rightarrow$ PM phase transitions are replaced by second-order phase transitions, in agreement with the results presented in Fig. \ref{POHZCWfixo}. The red solid circles indicate the positions of the tricritical points.
In addition, the dashed lines denote metamagnetic-like transitions. Such transitions can be observed in both AF phases, however for $k_{B}T=0.004$ eV, the transitions occur only for low values of $W$. The inset in Fig. \ref{WHZC}(b) highlights the region where the metamagnetic-like transitions occurs. For $k_{B}T=0.008$ eV the metamagnetic-like transitions no longer appear in the range of parameters considered.
\begin{figure}[!b]
    \centering
    \includegraphics[scale=1.0,angle=0]{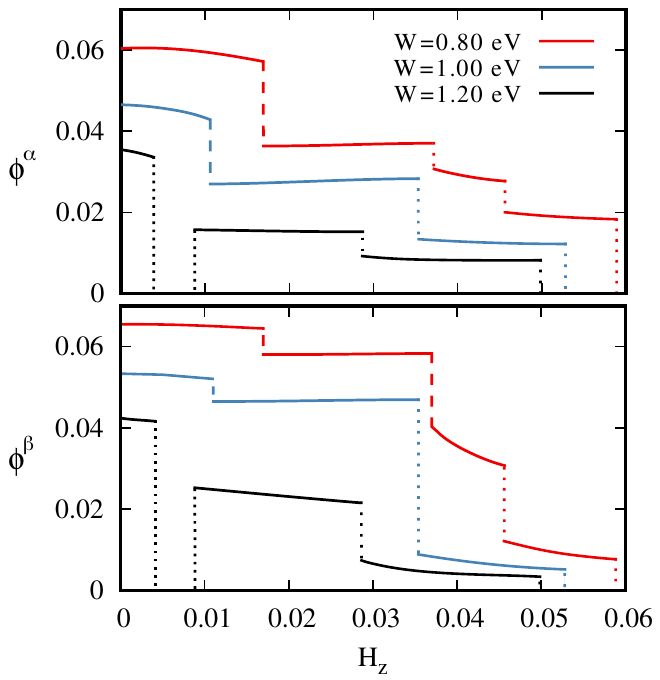}
    \caption{The Neel gaps $\phi^{\alpha}$ and $\phi^{\beta}$ for the tetragonal lattice as a function of $H_z$ for $T=0$ and different values for the band width $W$. The regions with dashed lines denotes the AF$_{1}\rightarrow$ AF$_{2}$ and AF$_{2}\rightarrow$ PM first-order transitions.}
    \label{POHZT}
\end{figure}

\begin{figure*}[!ht]
    \centering
    \includegraphics[scale=1.0,angle=-90]{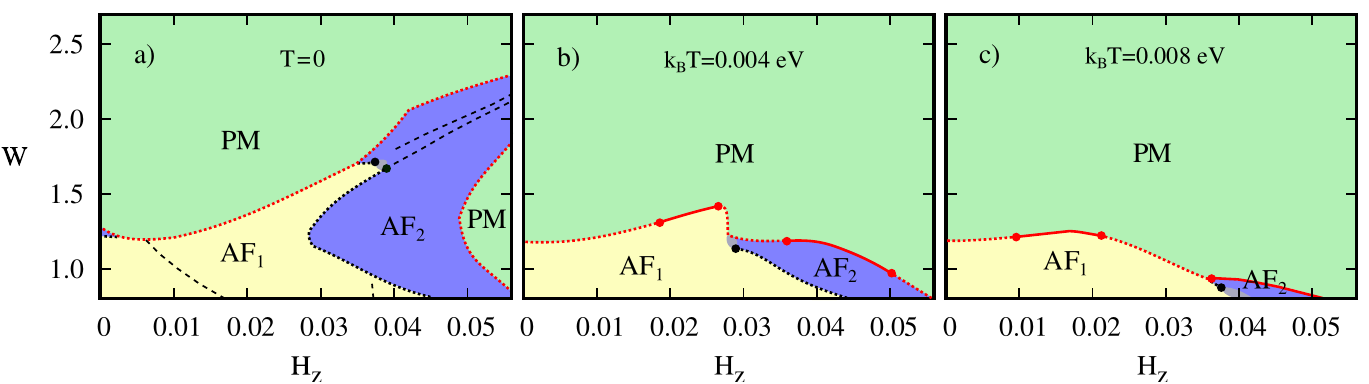}
    \caption{The  phase diagram  for a tetragonal lattice with the bandwidth $W$ versus $H_z$, for different temperatures. The solid and the dashed lines denote second-order and first-order
transition, respectively. The parameters of the dispersion relation are  $c/a=1.10$ and  $r=0.90$.}
    \label{WHZRT}
\end{figure*}

\begin{figure}[!hb]
    \centering
    \includegraphics[scale=0.85,angle=-90]{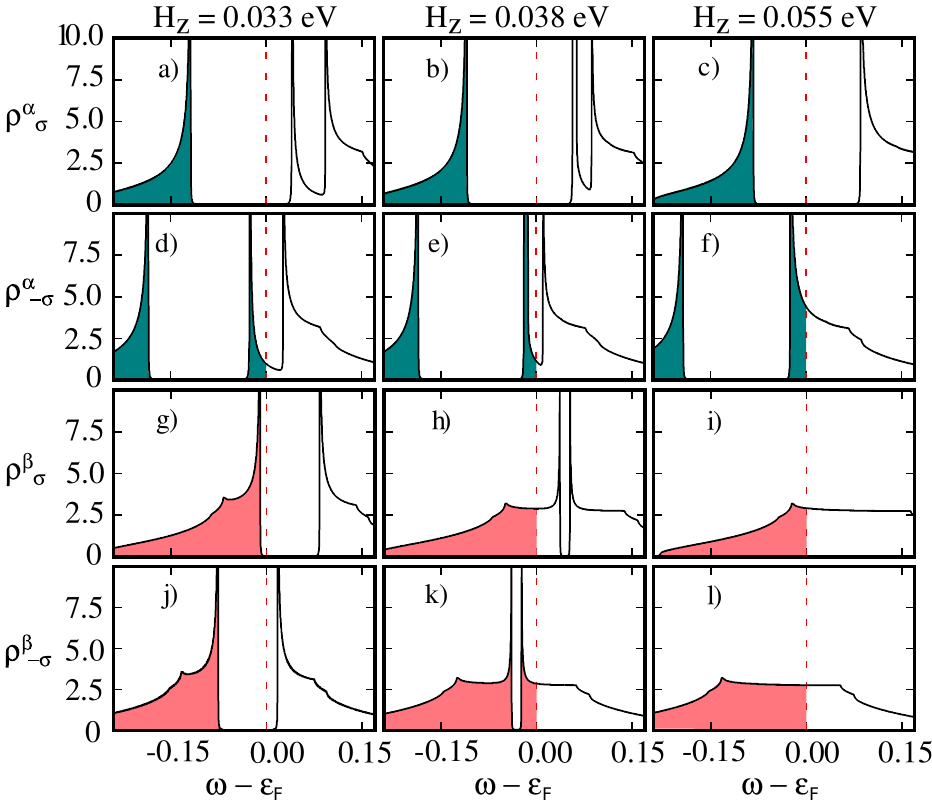}
    \caption{The $\alpha$ and $\beta$ partial densities of states for
W$=1.0$ eV, $T=0$, and different
values of $H_z$. The values of $H_z$ have been chosen in order to show the densities of states behavior inside each phase  of the diagram presented in the Fig. \ref{WHZRT}.    }
    \label{DOS_RTWfixo}
\end{figure}

In general, the discontinuities in the gaps as a function of 
$H_z$ (see Fig. \ref{POHZC}), are related to the position of the Fermi energy $\varepsilon_F$ relative to the gaps in the partial densities of states (DOS). In Fig. \ref{DOS_RCWfixo}, the DOS
associated with the sequence of transitions AF$_1$ $\rightarrow$AF$_2$ $\rightarrow$PM, are shown for $T=0$ and $W=1.20$ eV. The vertical dashed red lines indicate the position of the Fermi energy, for each case. The first and second columns of the panels shown the partial DOS for the AF$_1$ and AF$_2$ phases, respectively.  The third column shown the partial DOS in the PM phase of the system.
When 
$H_z$ increases from 0.030 eV to 0.032 eV, the Fermi energy moves out of the gap of the $\beta$-band partial DOS, $\rho^{\beta}_{\sigma}$, resulting in a discontinuity in the gap (see Fig. \ref{POHZC}) what gives rise to the AF$_1$ $\rightarrow$AF$_2$ phase transition. The positions of $\varepsilon_F$ in both cases, are shown in Figs. \ref{DOS_RCWfixo}(g) and \ref{DOS_RCWfixo}(h). In general, every time that $\varepsilon_F$ moves out of a gap in the DOS, due to an increase of either $H_z$ or $W$, the gaps change discontinuously (see Fig. \ref{POHZC}) and are accompanied by a phase transition or a metamagnetic-like transition.
\subsection{Tetragonal Lattice}

In this section we present results for the tetragonal lattice, i.e., $a\neq c$ and $r=t_{A,c}/t_{A,a}$. The crystalline symmetry lifts the degeneracy of the dispersion relations given in Eq. \ref{Ek}. In order to stay relatively close to the cubic lattice case, most of the results presented in this section were obtained using $c/a=1.10$, and $r=0.90$.

In Fig. \ref{POHZT}, it is seen that behavior of the Neel gaps for $W=0.80$ eV and $W=1.00$ eV,
is very similar to the behavior observed for the cubic lattice in Fig. \ref{POHZC}.
However, in the tetragonal case, a higher magnetic field is required to close the Neel gaps. For $W=1.20$ eV,
with an increase of $H_z$, the system leaves the phase AF$_1$ and enters in the PM phase in which the gaps are zero, for small values of $H_z$. If the magnetic field and therefore $H_z$ is further increased, the system reaches the AF$_1$ phase again.
When the magnetic field is increased to higher values, the system undergoes a first-order transition to AF$_{2}$ phase at $H_{z}\approx$ 0.028 eV and another first-order transition is found at $H_{z}\approx$ 0.05 eV where the system enters the PM phase.

The $W$ versus $H_z$ phase diagrams and their evolution with temperature, are shown in Fig. \ref{WHZRT}. For $T=0$, the region where the AF$_1$ phase occurs is similar to that of the cubic lattice. However, the AF$_2$ phase is concentrated in the region of higher magnetic field while in the cubic lattice the AF$_2$ phase also occurs for intermediate values of $H_z$. Indeed, the $\beta$ DOS for the tetragonal lattice is asymmetric relative to $\omega =0$ which results in the phase AF$_2$ being favored. The asymmetry can be seen, for example, in Fig. \ref{DOS_RTWfixo}(l). As in the case of the cubic lattice, two critical points (black solid circles) are present in the $T=0$ phase diagram.  For $k_{B}T=0.004$ eV, we observe the presence of four tricritial points (red solid circles) while in the cubic lattice the four tricritical points first occur at $k_{B}T=0.008$ eV. Furthermore, the critical point observed for $k_{B}T=0.004$ eV is still present for $k_{B}T=0.008$ eV. These facts indicate that the existence of critical and tricritical points is favored in the tetragonal lattice. On the other hand, the metamagnetic-like transitions represented by the dashed lines in Fig. \ref{WHZRT}(a), are less favored than in the cubic lattice.
\begin{figure}[!bth]
    \centering
    \includegraphics[scale=.80,angle=-90]{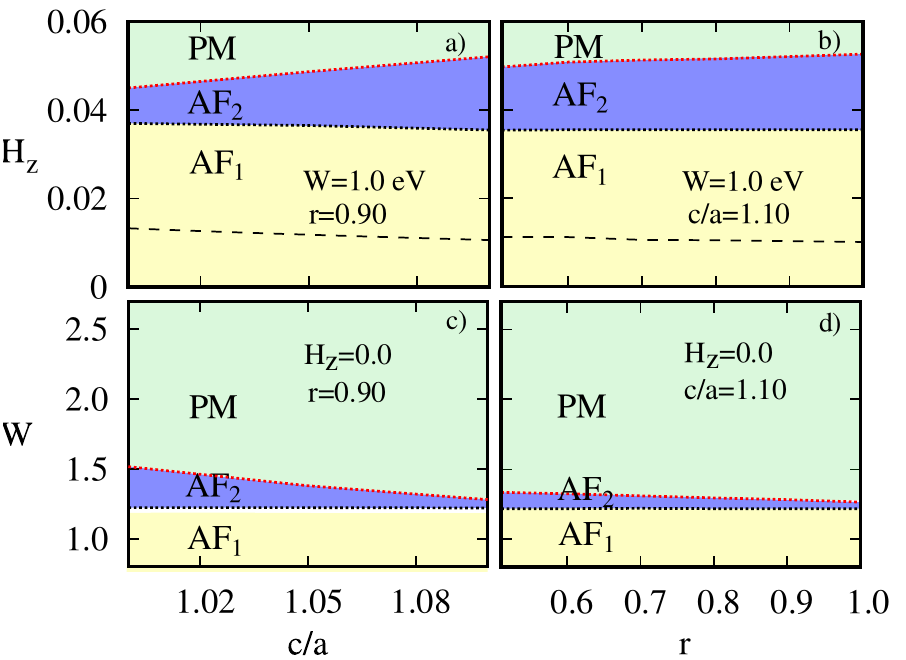}
    \caption{The phase diagram for the tetragonal lattice with $H_z$ and  $W$ versus  $c/a$ in the first column and versus $r$ in the second column.  All the transitions shown in the phase diagrams, have first order nature.}
    \label{HZ_Wvsr}
\end{figure}

The $\alpha$ and $\beta$ partial densities of states, $\rho^\alpha$ and $\rho^\beta$, are shown in Fig. \ref{DOS_RTWfixo} for $T=0$,
$W=1.0$ eV and different values of $H_z$. The values of $H_z$ in the AF$_1$ and AF$_2$ phases, were chosen in order to be close to the AF$_1$ $\rightarrow$ AF$_2$ phase transition.
The behavior of the Neel gaps for this set of parameters has been shown in Fig. \ref{POHZT}. By comparing the results in Fig. \ref{DOS_RTWfixo} with those for the cubic lattice shown in Fig. \ref{POHZC}, it is possible to see that the results are slightly different, mainly for $\rho_{-\sigma}^{\beta}$. 
For the phase AF$_1$ with $H_z$=0.033 eV, the position of the  Fermi energy, which is represented by the vertical dashed red line in Fig. \ref{DOS_RTWfixo}, is found inside the gap for both $\rho_{\sigma}^{\beta}$ and $\rho_{-\sigma}^{\beta}$ partial densities, while for the cubic lattice the Fermi energy is found inside the gap only for $\rho_{\sigma}^{\beta}$ (see Figs. \ref{DOS_RCWfixo}(g) and \ref{DOS_RCWfixo}(j)). This feature is related to the asymmetry of the partial DOS for the tetragonal lattice. For instance, in Fig. \ref{DOS_RTWfixo}(j), the area of the partial DOS $\rho_{-\sigma}^{\beta}$ below the Fermi energy (colored in red) is slightly larger than the area above the Fermi energy. In order to keep the total occupation of the bands constant, the Fermi energy has been moved to lower energies, i.e. into the gap of the DOS $\rho_{-\sigma}^{\beta}$, which results in a phase AF$_1$ that is less metallic when compared with the cubic case shown in Fig. \ref{DOS_RCWfixo}, for which the partial DOS $\rho^{\beta}$ is symmetric.

The results presented so far in this section have been obtained considering small deviations from the cubic lattice, for the parameters $c/a$ and $r$. Now, we investigate how the boundaries of the phases AF$_1$, AF$_2$ and PM behave when the parameters $c/a$ and $r$, are changed.
Fig. \ref{HZ_Wvsr}(a) exhibit the phase diagram with $H_z$ versus $c/a$, for fixed $W$ and $r$, while Fig. \ref{HZ_Wvsr}(c) exhibit the phase diagram with $W$ versus $c/a$, for $H_z$=0.0 and $r$ fixed. While the phase AF$_1$ is robust to the effects of $H_z$ and $W$ when $c/a$ is enhanced, the phase AF$_2$ is significantly affected by the increasing of $H_z$ or $W$, in this same situation.
Nevertheless, while the phase AF$_2$ is favored by the increasing of $c/a$ when $H_z$ is enhanced, the same phase is suppressed by the increasing of $c/a$, when $W$ is enhanced.
Such feature is related to the way that $H_z$ and $W$ affect the partial DOS, maily the $\rho_{\pm\sigma}^{\beta}$. While $H_z$ shifts $\rho_{\sigma}^{\beta}$ to lower energies and $\rho_{-\sigma}^{\beta}$ to higher energies, the main effect of $W$ is to increases the wide of the bands. Therefore,  the effects of $H_z$ combined with the asymmetry of the DOS $\rho_{\pm\sigma}^{\beta}$, relative to the gap (see Fig. \ref{DOS_RTWfixo}), are the main reasons for the features
present in the phase diagrams of Figs. \ref{HZ_Wvsr}(a) and
\ref{HZ_Wvsr}(c).
In Figs. \ref{HZ_Wvsr}(b) and
\ref{HZ_Wvsr}(d) it can be noted that the effect of to increasing $r$ keeping $c/a$ fixed, is similar to that one in which $r$ is kept fix while $c/a$ varies. However, the effects of varying $r$ are much less intense. The dashed lines in Figs.\ref{HZ_Wvsr}(b) and
\ref{HZ_Wvsr}(d) indicate the metamagnetic-like transitions.


\section{Conclusions}
\label{Conclusion}
In this work, we have investigated the effects of pressure and magnetic field $h_z$ on two distinct itinerant Neel phases using the underscreened Anderson Lattice Model (UALM).
The version of the UALM that we considered is composed by two narrow $f$-bands (of either $\alpha$ or $\beta$ character)
that hybridize with a single conduction band.
Besides the direct Coulomb interaction
between electrons in the same $5f$ band, we include a Hund’s rule
exchange interaction between electrons in the different bands \cite{Niklowitz210}.
We assume that application of pressure produces a variation of the bandwidth.
Moreover, given that the order parameter has an Ising-like
anisotropy, the magnetic field is considered parallel to this
anisotropy direction.
The Hund's rule exchange interaction $J$ couples the gaps $\phi^{\alpha}$ and $\phi^{\beta}$ of the different bands and give rise to two competing antiferromagnetic phases AF$_1$ and AF$_2$. The phase  AF$_1$ is characterized by $\phi^{\beta}>\phi^{\alpha}>0$ while in the phase AF$_2$ we have $\phi^{\alpha}>\phi^{\beta}>0$. The transition between these phases is first order. We analysed the UALM model for a cubic and a tetragonal lattice.

In order to investigate the effects of a magnetic field $h_z$ for different band widths $W$, we constructed $W$ versus $H_z$ phase diagrams for various temperatures. The $H_z$ and $h_z$ are related as in Eq. (\ref{eqHz}). The results show rich variety at $T=0$ for both lattices. In a previous work \cite{Magalhaes2020}, we investigated the effects of a
magnetic field oriented transverse to the $z$ axis with the same UALM model.
In that case, for a cubic lattice, the results showed that the increase of the transverse magnetic field suppresses the phase AF$_2$ while the phase AF$_1$ persists even at higher magnetic fields. In the present work, we find the opposite situation, i.e., the AF$_1$ phase is replaced by the AF$_2$ phase at higher magnetic fields
while the phase AF$_1$ occurs for lower values of $W$ and lower and intermediate values of $H_z$.
This dissemblance is related to the fact that the transverse field produces a spin-dependent momentum shift of the quasi-particles bands. On the other hand, the magnetic field $h_z$ splits the bands generating a spin-up and a spin-down sub-band \cite{Calegari2017}. An increase of $H_z$ shifts the spin-up and the spin-down sub-bands in opposite directions. The analysis of the partial density of states $\rho_{\pm\sigma}^{\alpha}$ and $\rho_{\pm\sigma}^{\beta}$ at $T=0$, helps us better understand why the field $h_z$ favors the phase AF$_2$. We demonstrated in Fig. \ref{DOS_RCWfixo} that in the AF$_1$ phase, the Fermi energy is inside the gaps of $\rho_{-\sigma}^{\alpha}$ and  $\rho_{\sigma}^{\beta}$, at least. On the other hand, if the Fermi energy is outside the gaps of $\rho_{\pm\sigma}^{\beta}$, but is still inside the gap of   $\rho_{-\sigma}^{\alpha}$, the system is found in the phase AF$_2$. Considering the fact that $H_z$ shifts the spin-up and the spin-down sub-bands in opposite directions, the configuration in which the Fermi energy is outside the gaps of both $\rho_{-\sigma}^{\beta}$ and $\rho_{\sigma}^{\beta}$, is favored when $H_z$ increases. Moreover, duo to the hybridization gap present in  $\rho_{\pm\sigma}^{\alpha}$, the gap
$\phi^{\alpha}$ is less affected by the magnetic field (see Fig. \ref{POHZC}), allowing the Fermi energy to remain inside the gap of $\rho_{-\sigma}^{\alpha}$, until higher values of $H_z$.
Indeed, these are the main reasons why the AF$_2$ phase is favored by the magnetic field $h_z$.

The phase diagram for the tetragonal lattice at $T=0$, shows that the phase AF$_2$ occurs at higher magnetic fields while in the cubic lattice it is present at moderate magnetic fields. Moreover, the phase AF$_2$ tends to be concentrated at lower values of $W$ and higher values of magnetic fields, in comparison with the cubic lattice case. The main reason for that, is an asymmetry present in the partial density of states of the tetragonal lattice, which is enhanced with the increasing of $c/a$, where $c$ and $a$ are the lattice parameters.
As a result, we observed that for a tetragonal lattice, the AF$_2$ phase is enhanced by the increasing of both $H_z$ and $c/a$. On the other hand, the increasing of $W$ and $c/a$ at the same time, is detrimental for the AF$_2$ phase. Otherwise, the phase AF$_1$ is much less sensitive to the variation of such parameters, including the temperature. In addition, another difference relative to the application of a transverse field \cite{Magalhaes2020}, is the presence of metamagnetic-like transitions which occurs in both AF phases under the application of a magnetic field $h_z$. We highlight that such phenomenology,  the metamagnetic-like transitions inside the antferromagnetic phases have been reported in some antiferromagnetic heavy fermions \cite{Ken} which also presents a competition between two distinct antiferromagnetic phases.



\section*{Acknowledgments}
The present study was partially supported by the brazilian agencies Conselho Nacional de Desenvolvimento Científico e Tecnológico (CNPq), Coordenação de Aperfeiçoamento de Pessoal de Nível Superior (CAPES), and Fundação de Amparo à pesquisa do Estado do RS (FAPERGS).
ACL acknowledges the hospitality of the Temple University, where part of this work was done.

\medskip

\begin{thebibliography}{00}


\bibitem{moore}
\href{https://doi.org/10.1103/RevModPhys.81.235}{
K. T. Moore and G. van der Laan, Rev. Mod. Phys. \textbf{81}, 235
(2009).}

\bibitem{Santini}
\href{https://doi.org/10.1080/000187399243419}{
 P.  Santini,  R.  Lémanski,  and  P.  Erdos, Adv. Phys. \textbf{48}, 537 (1999).}

 \bibitem{Miyake}
\href{https://doi.org/10.7566/JPSJ.88.063706}{
Atsushi Miyake,  Yusei Shimizu, Yoshiki J. Sato, Dexin Li, Ai Nakamura, Yoshiya Homma, Fuminori Honda, Jacques Flouquet, Masashi Tokunaga  and Dai Aoki, J. Phys. Soc. Jpn. \textbf{88}, 063706 (2019). }

\bibitem{Pfleiderer}
\href{https://doi.org/10.1103/RevModPhys.81.1551}{
C. Pfleiderer,Rev. Mod. Phys. \textbf{81}, 1551 (2009).}

\bibitem{Aoki}
\href{https://doi.org/10.7566/JPSJ.88.043702}{
Dai Aoki, Ai Nakamura, Fuminori Honda, DeXin Li, Yoshiya Homma, Yusei Shimizu, Yoshiki J. Sato, Georg Knebel, Jean-Pascal Brison, Alexandre Pourret, Daniel Braithwaite, Gerard Lapertot, Qun Niu, Michal Vališka, Hisatomo Harima, and Jacques Flouquet,
J. Phys. Soc. Jpn. \textbf{88}, 043702 (2019).}


\bibitem{Mydosh2011}
\href{https://doi.org/10.1103/RevModPhys.83.1301}{J. A. Mydosh and P. M. Openeer, Rev. Mod. Phys. {\bf 83}, 1301 (2011).}
%
\bibitem{Oppeneer2014}
\href{https://doi.org/10.1080/14786435.2014.916428}{J. A. Mydosh, P. M. Oppeneer, Philosophical Magazine {\bf 94} (32-33):3642-3662 (2014).}


\bibitem{Wolowiec}
\href{https://doi.org/10.1073/pnas.2026591118 }{
Wolowiec, Christian T. and Kanchanavatee, Noravee and Huang, Kevin and Ran, Sheng and Breindel, Alexander J. and Pouse, Naveen and Sasmal, Kalyan and Baumbach, Ryan E. and Chappell, Greta and Riseborough, Peter S. and Maple, M. Brian. Proceedings of the National Academy of Sciences \textbf{118}, 20 (2021).}

\bibitem{Calegari2017}
\href{https://doi.org/10.1038/s41535-017-0055-2}{E. J. Calegari, S. G. Magalhaes, P. S. Riseborough, npj Quant. Mat. {\bf 48}, 1 (2017).}

\bibitem{Magalhaes2020}
\href{https://doi.org/10.1103/PhysRevB.101.064407}{S. G. Magalhães, A. C. Lausmann, E. J. Calegari, and P. S. Riseborough, Phys. Rev. B, \textbf{101}, 064407 (2020).}

\bibitem{Faundez2021}
\href{https://doi.org/10.1088/1361-648X/abe476}{ J. Faundez, S.G . Magalhães, J. E. Calegari and P. S. Riseborough,  J. Phys.: Condens. Matter \textbf{33}, 295801 (2021). }

\bibitem{Valiska2018}
\href{https://doi.org/10.1103/PhysRevB.98.174439}{M. Vališka, H. Saito, T. Yanagisawa, C. Tabata, H. Amitsuka, K. Uhlirova, J. Prokleska, P. Proschele, J. Valenta, M. Misek, D. I. Gorbunov, J. Wosnitza, V. Sechovsky, Phys. Rev. B {\bf 98}, 174439 (2018).}
\bibitem{Correa2012}
\href{https://doi.org/10.1103/PhysRevLett.109.246405}{V. F. Correa, S. Francoual, M. Jaime, N. Harrison, T. P. Murray, E. C. Palm, S. W. Tolzer, A. H. Lacerda, P. A. Sharma. J. A. Mydosh,
Phys. Rev. Lett. {\bf 109}, 246405 (2012).}
\bibitem{Niklowitz210}
\href{https://doi.org/10.1103/PhysRevLett.104.106406}{P. G. Niklowitz, C. Pfleiderer, T. Keller, M. Vojta, Y. -K. Huang, J. A. Mydosh, Phys. Rev. Lett. {\bf 106}, 106406 (2010).}

\bibitem{Riseborough2012}
\href{https://doi.org/10.1103/PhysRevB.85.165116}{P. S. Riseborough, B. Coqblin, S. G. Magalhaes, Phys. Rev. B {\bf 85}, 165116 (2012).}


\bibitem{Perkins}
\href{https://doi.org/10.1103/PhysRevB.76.125101}{N. B. Perkins, M. D. Nunez-Regueiro, B. Coqblin, and J. R. Iglesias, Phys. Rev. B \textbf{76}, 125101 (2007).}

\bibitem{Thomas}
\href{https://doi.org/10.1103/PhysRevB.83.014415}{C. Thomas, A. S. da Rosa Simoes, J. R. Iglesias, C. Lacroix, N. B. Perkins, and B. Coqblin,  Phys. Rev. B \textbf{83}, 014415 (2011).}

\bibitem{Bernard}
\href{https://doi.org/10.1103/PhysRevB.92.094401}{B. Bernhard and C. Lacroix, Phys. Rev. B \textbf{92}, 094401 (2015).}

\bibitem{Shah2000}
\href{https://doi.org/10.1103/PhysRevB.61.564}{N. Shah, P. Chandra, P. Coleman, J. A. Mydosh,
Phys. Rev. B {\bf 61}, 564 (2000).}

\bibitem{Yuan1}
\href{https://doi.org/10.1088/2516-1075/ac0511}{
Xiao Yuan, Peter S Riseborough , E J Calegari and S G Magalhaes, Electronic Structure \textbf{3}, 2 (2021).}

\bibitem{Schoenes1981}
\href{https://doi.org/10.1016/0370-1573(80)90156-8}{J. Schoenes, Phys. Rep. {\bf 66}, 187 (1981).}
\bibitem{Havela1992}
\href{https://doi.org/10.1016/0921-4526(92)90087-9}{L. Havela, V. Sechovsky, F. R. de Boer, E. Bruck, H. Nakote, Physica B {\bf 177}, 159 (1992).}
\bibitem{Maskova2019}
\href{https://doi.org/10.1103/PhysRevB.99.064415}{S. Maskova, A. V. Andreev, Y. Skourski, S. Yasin, D. I. Gorbonov, S. Zherlitsyn, H. Nakotte, H. Kothapalli, F. Nasreen, C. Cupp, H. B. Cao, A. Kolomiets, L. Havela, Phys. Rev. B {\bf 99}, 064415 (2019).}
\bibitem{Ken}
\href{https://doi.org/10.1143/JPSJS.81SB.SB058}{Ken Iwakawa, Yusuke Hirose, Kentaro Enoki, Kiyohiro Sugiyama, Tetsuya Takeuchi, Fuminori Honda, Masayuki Hagiwara, Koichi Kindo, Takehito Nakano, Yasuo Nozue, Rikio Settai, and Yoshichika Ōnuki, J. Phys. Soc. Jpn.  \textbf{81}, (2012).}
\bibitem{Solhanek}
\href{https://doi.org/10.1103/PhysRevLett.95.026403}{
A. V. Silhanek, N. Harrison, C. D. Batista, M. Jaime, A. Lacerda, H. Amitsuka, and J. A. Mydosh
Phys. Rev. Lett. \textbf{95}, 026403 (2005).}

\bibitem{Mushnikov}
\href{https://doi.org/10.1103/PhysRevB.59.6877}{N. V. Mushnikov, T. Goto, K. Kamishima, H. Yamada, A. V. Andreev, Y. Shiokawa, A. Iwao, and V. Sechovsky, Phys. Rev. B \textbf{59}, 6877 (1999).}

\bibitem{Bruck1994}
\href{https://doi.org/10.1103/PhysRevB.49.8852}{E. Brück, H. Nakotte, F. R. de Boer, P. F. de Châtel, H. P. van der Meulen, J. J. M. Franse, A. A. Menovsky, N. H. Kim-Ngan, L. Havela, V. Sechovsky, J. A. A. J. Perenboom, N. C. Tuan, and J. Sebek, Phys. Rev. B {\bf 49}, 8852 (1994).}
%

\bibitem{Pospisil2018}
\href{https://doi.org/10.1103/PhysRevB.98.014430}{Jiří Pospíšil, Yoshinori Haga, Yoshimitsu Kohama, Atsushi Miyake, Shinsaku Kambe, Naoyuki Tateiwa, Michal Vališka, Petr Proschek, Jan Prokleška, Vladimír Sechovský, Masashi Tokunaga, Koichi Kindo, Akira Matsuo, and Etsuji Yamamoto, Phys. Rev. B {\bf 98}, 014430 (2018).}


\end{thebibliography}

\medskip

\end{document}